\documentclass[twocolumn,aps,showpacs,prb,tightenlines,amsmath,amssymb]{revtex4}
\usepackage{graphicx}%[dvips]
\usepackage{amssymb}
\usepackage{dcolumn}
\usepackage{amsmath}
\usepackage{bm}% bold math

\usepackage{colordvi}

\begin{document}

\title{Spin relaxation due to the Bir-Aronov-Pikus mechanism in
intrinsic and $p$-type GaAs quantum wells
from a fully microscopic approach}
\author{J.\ Zhou}
\author{M.\ W.\ Wu}
\thanks{Author to whom all correspondence should be addressed}
\email{mwwu@ustc.edu.cn.}
\affiliation{Hefei National Laboratory for Physical Sciences at
  Microscale, University of Science and Technology of China,
Hefei, Anhui, 230026, China}
\affiliation{Department of Physics, University of Science and
Technology of China, Hefei, Anhui, 230026, China}
\altaffiliation{Mailing address.}
\date{\today}

\begin{abstract}
We study the electron spin relaxation in intrinsic and  $p$-type
(001) GaAs quantum wells by constructing and numerically
solving the kinetic spin Bloch equations.
All the relevant scatterings
are explicitly included, especially the spin-flip
electron-heavy hole exchange scattering which leads to the Bir-Aronov-Pikus
spin relaxation. We show that, due to the neglection of the
nonlinear terms in the
electron-heavy hole exchange scattering in the
Fermi-golden-rule approach,
the spin relaxation due to the
Bir-Aronov-Pikus mechanism  is
greatly exaggerated at moderately high electron
density and low temperature in the literature.
We compare the spin relaxation time due to the
Bir-Aronov-Pikus mechanism with that due to the
D'yakonov-Perel' mechanism which is also calculated from the
kinetic spin Bloch equations with all the scatterings, especially
the  spin-conserving electron-electron and electron-heavy hole
scatterings, included.
We find that, in intrinsic quantum wells, the effect from the
Bir-Aronov-Pikus mechanism is much smaller than that from the D'yakonov-Perel'
mechanism at low temperature, and
it is smaller by no more than one order of magnitude at high temperature.
In $p$-type quantum wells, the spin relaxation due to the
Bir-Aronov-Pikus mechanism is also much smaller
than the one  due to the D'yakonov-Perel'  mechanism
at low temperature and
becomes  comparable to each other at higher temperature when the hole
density and the width of the quantum well are large enough.
We claim that unlike in the bulk samples, the Bir-Aronov-Pikus mechanism
hardly dominates the spin relaxation in two-dimensional samples.

\end{abstract}


\maketitle

\section{INTRODUCTION}

Much attention has been given to semiconductor spintronics
both theoretically and experimentally due to
great prospect of the potential applications.\cite{meier,prinz,Fabian}
The study of the spin relaxation/dephasing (R/D) in semiconductors
contains rich physics and is of great importance for the device
application. Three spin R/D  mechanisms
have long been proposed in Zinc-blend semiconductors, i.e.,
the Elliott-Yafet  (EY) mechanism,\cite{EY} caused by
the spin-flip electron-impurity scattering due to
the spin-orbit coupling;
the D'yakonov-Perel' (DP) mechanism\cite{DP} which is due to the
momentum-dependent spin splitting
in crystal without a center of symmetry;
and the Bir-Aronov-Pikus (BAP)
mechanism\cite{bap} which originates
from the spin-flip electron-hole exchange interaction.
Previous researches have shown that, in bulk systems, the EY
mechanism is important in
narrow-band-gap and high impurity semiconductors; the DP mechanism is
dominant in $n$-type semiconductors; and the BAP mechanism can
have significant effect in $p$-doped semiconductors.\cite{hun,aronov,zerr}
It is  known that, in heavily $p$-doped bulk samples,
the BAP mechanism is dominant at low
temperature whereas the DP mechanism is dominant at high temperature with
the crossover temperature determined by the doping level. In bulk samples
with low hole density, the BAP mechanism has been
shown to be irrelevant.\cite{hun,aronov,zerr}
In addition, the hyperfine interaction induced spin relaxation is
 another possible mechanism.\cite{pershin}

In contrast to the bulk systems,
the relative importance of the BAP and DP mechanisms
for the electron spin R/D
in two dimensional (2D)
systems, especially in $p$-type 2D systems, is still not very clear,
sometimes even confusing. In Ref.\ [\onlinecite{wagner}], extremely long
spin relaxation time (SRT), which is two orders of magnitude longer
than that in the bulk sample with corresponding
acceptor concentrations, was reported by Wagner {\em et al.} in
$p$-type GaAs quantum wells (QWs).
The authors argued that the BAP mechanism is dominant at low
temperature. However in Ref.\ [\onlinecite{sham}], the SRT in
$p$-type QWs was
reported to be a factor of 4 shorter than that in comparably bulk
GaAs by Damen {\em et al.} at low temperature.
The authors also referred the BAP mechanism
as a cause for the decrease of the SRT.
Hence, two opposite experimental results arrive at the same conclusions
 regarding the importance of the BAP mechanism.
Moreover, Gotoh {\em et al.} further  pointed out that the BAP
mechanism should not be ignored even at room temperature.\cite{gotoh}
They investigated the electric
field dependence of SRT and found that the SRT decreases with the
increase of the bias.
They concluded that the decrease is from the BAP mechanism
as the SRT due to the DP mechanism does not change with electric field.
Actually, they overlooked the fact
 that the Rashba spin-orbit coupling\cite{Rashba} can also
lead to the spin R/D due to the DP mechanism.
 Therefore, we believe that the decrease of SRT in their
experiment cannot be a proof of the importance of the BAP mechanism.
Very recently it was shown that the SRT at room temperature
can be increased at the $(100)$ GaAs surface due to the relatively lower
concentration of holes at the surface and the mechanism for the SRT
was referred to as the BAP mechanism.\cite{schneider}
Theoretically, Maialle\cite{maialle}
pointed out that the effect of the BAP mechanism in 2D systems
 is a little smaller than that of
the DP mechanism  at zero temperature by using the
Fermi golden rule to calculate the SRT in which the elastic
scattering approximation was applied and consequently the nonlinear
terms of the electron-hole Coulomb scattering were neglected.
The SRT due to the DP mechanism ($\tau_{\mbox{\tiny DP}}$) was also calculated
by using the single particle approach.\cite{meier,DP}
The author compared $\tau_{\mbox{\tiny DP}}$ and
$\tau_{\mbox{\tiny BAP}}$ for different
electron momentums (kinetic energies), and showed that these two SRTs
have nearly the same order of magnitude in heavily doped QWs.
However, $\tau_{\mbox{\tiny DP}}$ calculated in Refs.\
[\onlinecite{maialle}] is quite cursory, because,
under the framework of single particle
theory the carrier-carrier Coulomb scattering, which is
very important to spin R/D,\cite{wu1,wu7,wu8,wu12,ivchenko} is not included.
Also the counter effect
of the scattering to the spin R/D is also not fully
accounted.\cite{wu7,wu8,wu12,wu9,lue}
Moreover, it is also important to calculate the spin-flip
electron-hole exchange scattering explicitly in order to find
out the effect of the nonlinear terms ignored in the
Fermi golden rule approach by
Maialle {\em et al.}\cite{maialle,maialle1,maialle2}
We also want to find out the temperature dependence of the
relative importance of both mechanisms in 2D systems, which to the
best of our knowledge is still absent in the literature..

In order to accurately investigate the relative importance of the DP
and the BAP mechanisms beyond the single-particle  Fermi
golden rule approach, we use the fully microscopic approach
established by Wu {\em et al.}\cite{wu}
by constructing and numerically solving the kinetic spin Bloch
equations.\cite{wu2,wu1,wu7,wu8,wu9,lue,wu12}
In this approach, all the corresponding
scatterings such as the electron-acoustic (AC) phonon,
electron-longitudinal optical (LO) phonon,
electron-nonmagnetic impurity, and electron-electron Coulomb
scatterings are explicitly included. The results/predictions obtained
from this approach are in very good agreement with varies
experiments.\cite{wu12,wu11,exp1,exp2}
It was previously pointed out
that, in the presence of inhomogeneous broadening, any type of
scattering, including the Coulomb scattering, can give rise
to the spin R/D.\cite{wu1,wu7,wu8,lue,wu12}
In this paper, in addition to the all the above mentioned
scatterings in $n$-type QWs as considered in Ref.\
[\onlinecite{wu12}], we further add the
spin-conserving and spin-flip electron-heavy hole Coulomb scatterings,
both contributing to the DP mechanism
and the latter further leading to the spin R/D due to the BAP mechanism.
By solving the kinetic spin Bloch equations self-consistently,
we obtain the SRT from the BAP mechanism from a fully microscopic
fashion. We further
investigate the relative importance of the BAP and DP
mechanisms in 2D systems.

This paper is organized as follows. In Sec.\ II, we construct the
kinetic spin Bloch equations and present the scattering terms from
the spin-conserving and spin-flip electron-hole
Coulomb scatterings. We also discuss the
SRTs due to the BAP mechanism from different approaches.
Then we present our numerical results in Sec.\ III. We study the
SRT due to both the DP and the BAP mechanisms
under various conditions such as temperatures,
electron/hole densities, impurity densities,
and well widths. We conclude in Sec.\ VI.

\section{KINETIC SPIN BLOCH EQUATIONS}

We construct the kinetic spin Bloch equations in intrinsic and
$p$-type (001) GaAs QWs by using the
nonequilibrium Green's function method:\cite{haug}
\begin{equation}
\dot{\rho}_{{\bf k},\sigma \sigma^{\prime}}=\dot{\rho}_{{\bf k},
\sigma \sigma^{\prime}}|_{\mbox{coh}}
+\dot{\rho}_{{\bf k},\sigma \sigma^{\prime}}|_{\mbox{scatt}}\ ,
\label{bloch}
\end{equation}
with $\rho_{{\bf k},\sigma \sigma^{\prime}}$ representing the
single particle density matrix elements.
The diagonal and off-diagonal
elements of $\rho_{{\bf k},\sigma \sigma^{\prime}}$
give the electron distribution functions $f_{{\bf k}\sigma}$ and the spin
coherence $\rho_{{\bf k},\sigma-\sigma}$, respectively.
The coherent terms $\dot{\rho}_{k,\sigma \sigma^{\prime}}|_{\mbox{coh}}$
describe the precession
of the electron spin due to the effective magnetic field from
the Dresselhaus term\cite{dress}
$\mathbf{\Omega}({\bf k})$ and the Hartree-Fock Coulomb interaction.
The expression of the coherent terms can be found in
Appendix A (and also Ref.\ [\onlinecite{wu7}]).
The Dresselhaus term can be written as:\cite{dp2}
\begin{eqnarray}
\label{omegax}
\Omega_x({\bf k})&=&\gamma k_x(k_y^2-\langle k_z^2\rangle), \\
\Omega_y({\bf k})&=&\gamma k_y(\langle k_z^2\rangle-k_x^2),  \\
\label{omegaz}
\Omega_z({\bf k})&=&0\ ,
\end{eqnarray}
in which $\langle k_z^2\rangle$ represents the average of the operator
$-(\partial/\partial z)^2$ over the electronic state of the lowest
subband,\cite{wu12}
and $\gamma$ is the spin splitting parameter\cite{meier} which is
chosen to be $11.4$\ eV$\cdot$\AA$^3$ all through the paper.\cite{comment}
$\dot{\rho}_{k,\sigma \sigma^{\prime}}|_{\mbox{scatt}}$ in Eq.\ (\ref{bloch})
denote the electron-LO-phonon, electron-AC-phonon,
electron-nonmagnetic impurity,
and the electron-electron Coulomb scatterings
whose expressions are given in detail in  Appendix A
(see also Refs.\ [\onlinecite{wu7,wu8,wu12}]).
All these scattering are
calculated {\em explicitly} without any relaxation time
approximation. Moreover, we further include the spin-conserving and spin-flip
electron-heavy hole scatterings as what follows.

The Halmitonian of electron-heavy hole interaction is given by
\begin{equation}
H_{\mbox{\small eh}}=\sum_{{\bf k},
 {\bf  k}^{\prime},{\bf q},\sigma=\pm1,\sigma^\prime=\pm1}V_{eh,q}c_{{\bf
    k}+{\bf q},\frac{\sigma}{2}}^{\dagger}c_{{\bf k},\frac{\sigma}{2}}b_{{\bf
    k}^\prime-{\bf q},\frac{3\sigma^\prime}{2}}^{\dagger}b_{{\bf k^{\prime}},
\frac{3\sigma^\prime}{2}} \ ,
\end{equation}
where $c$ ($c^{\dagger}$) and $b$ ($b^{\dagger}$) are the annihilation
(creation) operators of electrons in conduction  (heavy-hole
valence) band respectively. We denote $\sigma$ ($\sigma^{\prime}$)
to be $\pm 1$ throughout the paper.
The screend Coulomb potential under the random-phase
approximation reads\cite{haug}
\begin{equation}
V_{eh,q}=\frac{\sum_{q_z}v_{Q}f_{eh}(q_z)}{\epsilon ({\bf q})} \ ,
\end{equation}
with the bare Coulomb potential $v_{Q}=4\pi e^2/Q^2$ and
\begin{eqnarray}
\epsilon({\bf q})&=&1-\sum_{q_{z}}v_{Q}f_e(q_{z})\sum_{{\bf k},\sigma}
\frac{f_{{\bf k}+{\bf q},\sigma}
-f_{{\bf k},\sigma}}{\varepsilon^{e}_{\bf k+q}-\varepsilon^e_{\bf k}}
\nonumber \\
&&\mbox{}-\sum_{q_{z}}v_{Q}f_h(q_{z})\sum_{{\bf k^{\prime}},\sigma}\frac{f^h_{{\bf k^{\prime}}+{\bf q},\sigma}
-f^h_{{\bf k^{\prime}},\sigma}}{\varepsilon^{h}_{\bf
    k^{\prime+q}}-\varepsilon^h_{\bf k^{\prime}}}
\end{eqnarray}
is the electron-hole plasma screening.\cite{comment1}
In these equations $Q^2={\bf q}^2+q_z^2$ and $f^h_{{\bf k},\sigma}$
($f_{{\bf k},\sigma}$)
denotes the heavy hole (electron) distribution function
with spin $\frac{3}{2} \sigma$ ($\frac{1}{2} \sigma$).
The form factors can be written as:
\begin{eqnarray}
\label{forme}
&&\hspace{-0.5cm}f_{e}(q_z)=\int dz dz^{\prime}\xi_{c}(z) \xi_c(z^{\prime}) e^{i q_z(z-z^{\prime})}
\xi_c(z^{\prime})\xi_{c}(z) \ ,\\
\label{formh}
&&\hspace{-0.5cm}
f_{h}(q_z)=\int dz dz^{\prime} \eta_{h}(z) \eta_h(z^{\prime})
 e^{i q_z(z-z^{\prime})}  \eta_h(z^{\prime})\eta_{h}(z) \ ,\\
\label{form}
&&\hspace{-0.5cm}f_{eh}(q_z)=\int dz dz^{\prime} \xi_{c}(z) \eta_h(z^{\prime})
e^{i q_z(z-z^{\prime})}\eta_h(z^{\prime})\xi_{c}(z) \ ,
\end{eqnarray}
where $\xi_{c}(z)$ ($\eta_{h}(z)$) is the envelope function of the electron
(heavy hole) along the growth direction $z$.\cite{wu12}
The scattering term of this spin-conserving  electron-hole
Coulomb  scattering
can be written as:
\begin{widetext}
\begin{eqnarray}
\left.\frac{\partial f_{{\bf k},\sigma}}{\partial t}\right|_{\mbox{\small
eh}}&=&-2\pi\sum _{{\bf k^{\prime},q},\sigma^{\prime}}
\delta(\varepsilon_{\bf k-q}^e-\varepsilon_{\bf k}^e
+\varepsilon_{\bf k^{\prime}}^h-\varepsilon_{\bf k^{\prime}-q}^h)
V_{eh,q}^2\Bigl\{ (1-f_{{\bf k^{\prime}},\sigma^{\prime}}^h)f_{{\bf k^{\prime}-q},\sigma^{\prime}}^h\nonumber\\
&&\times [f_{{\bf k},\sigma}(1-f_{{\bf k-q},\sigma})-\mbox{Re}(\rho_{\bf k}\rho_{\bf k-q}^{\ast})]
-f_{{\bf k^{\prime}},\sigma^{\prime}}^h(1-f_{{\bf k^{\prime}-q},\sigma^{\prime}}^h)
[f_{{\bf k-q},\sigma}(1-f_{{\bf k},\sigma})-\mbox{Re}(\rho_{\bf k}\rho_{\bf k-q}^{\ast})] \Bigr\},\\
\left. \frac{\partial \rho_{\bf k}}{\partial t}\right |_{\mbox{\small
eh}}&=&-\pi\sum_{{\bf k^{\prime},q},\sigma,\sigma^{\prime}}
  \delta(\varepsilon_{\bf k-q}^e-\varepsilon_{\bf k}^e
+\varepsilon_{\bf k^{\prime}}^h-\varepsilon_{\bf k^{\prime}-q}^h)
V_{eh,q}^2\Bigl\{(1-f_{{\bf k^{\prime}},\sigma^{\prime}}^h)f_{{\bf k^{\prime}-q},\sigma^{\prime}}^h\nonumber\\
&&\times [(1-f_{{\bf k-q},\sigma})\rho_{\bf k}-f_{{\bf k},\sigma}\rho_{\bf k-q}]
+f_{{\bf k^{\prime}},\sigma^{\prime}}^h(1-f_{{\bf k^{\prime}-q},\sigma^{\prime}}^h)
  [f_{{\bf k-q},\sigma}\rho_{\bf k}-(1-f_{{\bf k},\sigma})\rho_{\bf k-q}] \Bigr\}\ ,
\end{eqnarray}
\end{widetext}
where $\rho_{\bf k}\equiv\rho_{{\bf k},\frac{1}{2}-\frac{1}{2}}
\equiv \rho_{{\bf k},-\frac{1}{2}\frac{1}{2}}^\ast$.
This spin-conserving scattering only enhances the total
scattering strength moderately and contributes to the
spin R/D due to the DP mechanism.

The Hamiltonian of the spin-flip electron-heavy hole exchange interaction reads
\begin{equation}
H_{\mbox{\small BAP}}=\sum_{{\bf k,
    k^{\prime},q},\sigma}M_{\sigma}({\bf k},{\bf k}^{\prime})c_{{\bf
    k}+{\bf q},\frac{\sigma}{2}}^{\dagger}b_{{\bf k}^{\prime}-{\bf q},
-\frac{3\sigma}{2}}^{\dagger}
c_{{\bf k},-\frac{\sigma}{2}}b_{{\bf k}^{\prime},\frac{3\sigma}{2}}\  .
\end{equation}
The matrix elements in the Hamiltonian are given by\cite{maialle2}
\begin{equation}
M_{\sigma}({\bf k},{\bf k}^{\prime})=\frac{3}{8}\frac{\Delta
E_{LT}}{|\phi_{3D}(0)|^2}\sum_{q_z}\frac{f_{ex}(q_z)(k^2_{\sigma}+k^{\prime
    2}_{\sigma})}{q_z^2+|{\bf k}+{\bf k}^{\prime}|^2} \ ,
\end{equation}
where $\Delta E_{LT}$ is the longitudinal-transverse splitting in
bulk, $|\phi_{3D}(0)|^2={1}/{(\pi a_0^3)}$ is the 3D
exciton state at zero relative distance, and $k_{\sigma}=k_{x}+i\sigma
k_{y}$.  For GaAs, $\Delta E_{LT}=0.08$\ meV and $a_0=146.1$\ \AA\
 respectively.\cite{ekardt}
The form factor can be written as:
\begin{equation}
f_{ex}(q_z)=\int dz dz^{\prime} \xi_{c}(z^{\prime}) \eta_h(z^{\prime})
e^{i q_z(z-z^{\prime})}  \eta_h(z)\xi_{c}(z) \ .
\label{form1}
\end{equation}
The scattering term from this Hamiltonian reads
\begin{widetext}
\begin{eqnarray}
\left.\frac{\partial f_{{\bf k},\sigma}}{\partial t}\right|_{\mbox{\small
BAP}}&=&-2\pi\sum _{{\bf k^{\prime},q}}
\delta(\varepsilon_{\bf k-q}^e-\varepsilon_{\bf k}^e
+\varepsilon_{\bf k^{\prime}}^h-\varepsilon_{\bf k^{\prime}-q}^h)
M_{\sigma}({\bf k-q,k^{\prime}})M_{-\sigma}({\bf k,k^{\prime}-q})
\nonumber\\
&&\times [(1-f_{{\bf k^{\prime}},\sigma}^h)f_{{\bf k^{\prime}-q},-\sigma}^h
f_{{\bf k},\sigma}(1-f_{{\bf k-q},-\sigma})
-f_{{\bf k^{\prime}},\sigma}^h(1-f_{{\bf k^{\prime}-q},-\sigma}^h)
(1-f_{{\bf k},\sigma})f_{{\bf k-q},-\sigma}]
\label{BAPscat1} ,\\
\left. \frac{\partial \rho_{\bf k}}{\partial t}\right |_{\mbox{\small
BAP}}&=&-\pi\sum_{{\bf k^{\prime},q},\sigma}
  \delta(\varepsilon_{\bf k-q}^e-\varepsilon_{\bf k}^e
+\varepsilon_{\bf k^{\prime}}^h-\varepsilon_{\bf k^{\prime}-q}^h)
M_{\sigma}({\bf k-q,k^{\prime}})M_{-\sigma}({\bf k,k^{\prime}-q})\nonumber\\
&&\times [(1-f_{{\bf k^{\prime}},\sigma}^h)f_{{\bf k^{\prime}-q},-\sigma}^h
  (1-f_{{\bf k-q},-\sigma})\rho_{\bf k}+
  f_{{\bf k^{\prime}},\sigma}^h(1-f_{{\bf k^{\prime}-q},-\sigma}^h)
  f_{{\bf k-q},\sigma}\rho_{\bf k}] \ .
\label{BAPscat}
\end{eqnarray}
\end{widetext}
If we denote ${\bf K}={\bf k}+{\bf k^{\prime}}$ as
the center-of-mass momentum of the electron-hole pair,
the product of the matrix elements in Eqs.\ (\ref{BAPscat1}) and
(\ref{BAPscat}) can be reduced to:
\begin{eqnarray}
&&|M({\bf K-q})|^2=M_{\sigma}({\bf k-q,k^{\prime}})M_{-\sigma}({\mathbf
k,{\mathbf k}^{\prime}-{\mathbf q}}) \nonumber \\
&&=\frac{9\Delta
E_{LT}^2}{16|\phi_{3D}(0)|^4}[\sum_{q_z}\frac{f_{ex}(q_z)({\bf K-q})^2}
{q_z^2+({\bf K-q})^2}]^2 \ .
\label{element}
\end{eqnarray}
It is noted that the spin R/D of the photo-excited holes is
very fast\cite{lue} and the electron-hole recombination is very slow compared
to the electron spin R/D. Therefore, we take the hole distribution
in equilibrium Fermi distribution and $f^h_{{\bf k}\sigma}=f^h_{{\bf k}-\sigma}
\equiv f^h_{\bf k}$. Further, by subtracting
$\left.\frac{\partial f_{{\bf k},-1}}{\partial t}\right|_{\mbox{\small
BAP}}$ from  $\left.\frac{\partial f_{{\bf k},+1}}{\partial t}\right|_{\mbox{\small
BAP}}$ in Eq.\ (\ref{BAPscat1}), one obtains:
\begin{widetext}
\begin{eqnarray}
\left.\frac{\partial \Delta f_{\bf k}}{\partial t}\right|_{\mbox{\small
BAP}}&=&\left.\frac{\partial (f_{{\bf k},+1}-f_{{\bf k},-1})}
{\partial t}\right|_{\mbox{\small
BAP}}=-2\pi\sum _{{\bf k^{\prime},q}}
\delta(\varepsilon_{\bf k-q}^e-\varepsilon_{\bf k}^e
+\varepsilon_{\bf k^{\prime}}^h-\varepsilon_{\bf k^{\prime}-q}^h)
|M({\bf K-q})|^2
\Bigl\{\Delta f_{\bf k}[(1-f^h_{\bf
    k^{\prime}})f^h_{\bf k^{\prime}-q}\nonumber\\
&&\hspace{-1.5cm}+\frac{1}{2}(f^h_{\bf k^{\prime}}-f^h_{\bf
    k^{\prime}-q})(f_{{\bf k-q},+1}+f_{{\bf k-q},-1})]+\Delta f_{\bf k-q}[f^h_{\bf
    k^{\prime}}(1-f^h_{\bf k^{\prime}-q})-\frac{1}{2}(f^h_{\bf k^{\prime}}-f^h_{\bf
   k^{\prime}-q})(f_{{\bf k},+1}+f_{{\bf k},-1})] \Bigr\} \ .
\label{del}
\end{eqnarray}
\end{widetext}
%It is clear from above equation that the SRT due to the BAP mechanism is
%inversely proportional to the matrix element of the spin-flip
%electron-heavy hole exchange scattering.
In above equation, the terms $\Delta f_{\bf k}[(1-f^h_{\bf
    k^{\prime}})f^h_{\bf k^{\prime}-q}+\frac{1}{2}(f^h_{\bf k^{\prime}}-f^h_{\bf
    k^{\prime}-q})(f_{{\bf k-q},+1})+f_{{\bf k-q},-1})]$
describe the forward
scattering and correspondingly
the terms $\Delta f_{\bf k-q}[f^h_{\bf
    k^{\prime}}(1-f^h_{\bf k^{\prime}-q})-\frac{1}{2}(f^h_{\bf k^{\prime}}-f^h_{\bf
   k^{\prime}-q})(f_{{\bf k},+1}+f_{{\bf k},-1})]$ describe the backward
scattering. The SRT due to the BAP mechanism from the Fermi golden
rule\cite{maialle} can be recovered from Eq.\ (\ref{del})
by applying the elastic scattering approximation:
$\varepsilon_{\bf k-q}^e \approx \varepsilon_{\bf k}^e$ and $\varepsilon_{\bf
   k^{\prime}}^h \approx \varepsilon_{\bf k^{\prime}-q}^h$.
Under this approximation, the nonlinear
terms (in the sense of the electron distribution function)
$\frac{1}{2}\Delta f_{\bf k}(f^h_{\bf k^{\prime}}-f^h_{\bf
   k^{\prime}-q})(f_{{\bf k-q},+1}+f_{{\bf k-q},-1})$ in the forward
scattering and $\frac{1}{2}\Delta f_{\bf k-q}(f^h_{\bf k^{\prime}}-f^h_{\bf
  k^{\prime}-q})(f_{{\bf k},+1}+f_{{\bf k},-1})$ in the backward scattering tend to zero.
In the remaining linear terms, $\Delta f_{\bf k}=-\frac{\partial
  f_{0{\bf k}}}{\partial \varepsilon_{\bf k}} (\phi_{1/2}-\phi_{-1/2})
\approx\Delta f_{\bf k-q}$
with $f_{0{\bf k}}=\frac{1}{e^{\beta(\varepsilon_{\bf k}
-\mu)}+1}$ by choosing
 $f_{{\bf k},\sigma}=\frac{1}{e^{\beta(\varepsilon_{\bf k}
-\mu-\phi_{\sigma})}+1}$.
Therefore, one recovers the SRT due to the BAP mechanism from the
Fermi golden rule approach:\cite{maialle}
\begin{eqnarray}
\frac{1}{2\tau^1_{\mbox{\tiny BAP}}({\bf k})}&=&2\pi\sum _{{\bf k^{\prime},q}}
\delta(\varepsilon_{\bf k-q}^e-\varepsilon_{\bf k}^e
+\varepsilon_{\bf k^{\prime}}^h-\varepsilon_{\bf k^{\prime}-q}^h) \nonumber \\
&&\mbox{}\times|M({\bf K-q})|^2[(1-f^h_{\bf k^{\prime}})f^h_{\bf k^{\prime}-q}] \ .
\label{tauBAP}
\end{eqnarray}
In the next section, we will discuss the applicability of above equation
which relies on the elastic scattering approximation.

In this work, we do not use the SRTs from the single-particle approach
for both the BAP and DP mechanisms. Instead, we
solve the  kinetic spin Bloch equations self-consistently with all the
scattering explicitly included. The detail of the
numerical scheme is given  in Refs.\ [\onlinecite{wu8,wu12}]. The
spin relaxation and dephasing times can be obtained from  the
temporal evolutions of the electron distribution functions $f_{{\bf k},\sigma}$
and the  spin coherence $\rho_{{\bf k},\sigma-\sigma}$
respectively.\cite{wu2,tts}
We will show that the SRT due to the BAP mechanism obtained from the
kinetic spin Bloch approach can give markedly different results compared to
the one calculated from Eq.\ (\ref{tauBAP}) by using the
elastic scattering approximation, similar to the situation of the SRT
due to the DP mechanism which has been discussed in great detail in our
previous works.\cite{wu7,wu8,wu12}

\section{Numerical RESULTS and Analysis}

The SRTs calculated from the kinetic spin Bloch equations are
plotted in Figs.\ 1 to 6. In these figures, the solid curves
represent the SRTs due to the BAP mechanism ($\tau_{\mbox{\tiny BAP}}$)
 which are calculated from the kinetic spin Bloch equations
by setting  the DP term ${\bf \Omega}({\bf k})=0$;
the dashed curves are the SRTs due to
the DP mechanism ($\tau_{\mbox{\tiny DP}}$)
which are calculated by setting
$\partial \rho_{{\bf k},\sigma\sigma^\prime}/\partial t|_{\mbox{\small
BAP}}=0$;
and the dash-dotted curves represent the total SRTs
($\tau_{\mbox{\tiny total}}$) obtained from Eq.\ (\ref{bloch}) with all the
terms explicitly included. We always use different color and width
of curves for different conditions.

\begin{figure}[htb]
\begin{center}
\includegraphics[height=5.5cm]{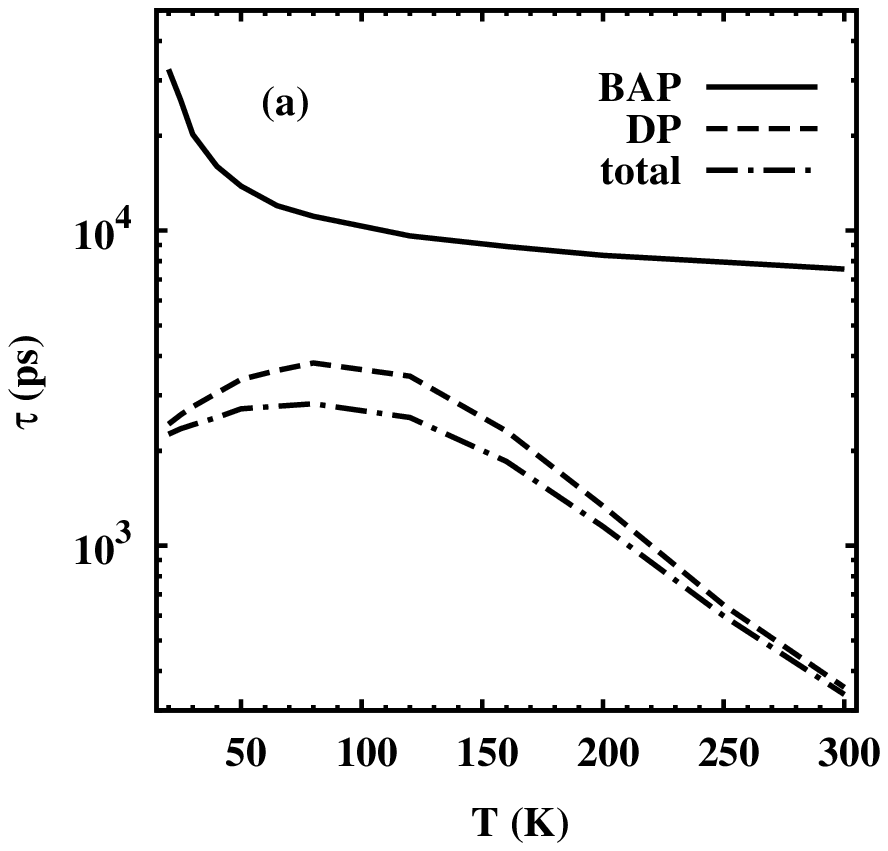}
\includegraphics[height=5.5cm]{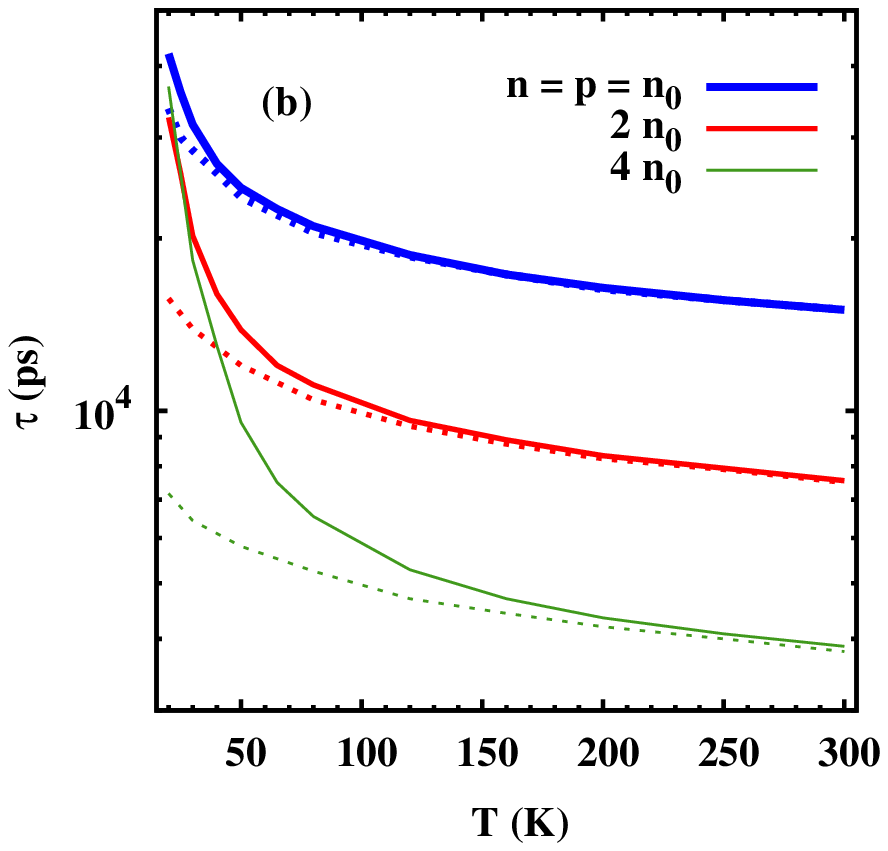}
\end{center}
\caption{(a) SRT  due to the BAP (solid curve) and
DP (dashed curve) mechanisms and
the total SRT (dash-dotted curve) {\em vs.} temperature $T$ in
  intrinsic QW when $a =20$\ nm, electron and hole densities
$n = p = 2 n_0$,  and impurity density $n_i = n$.
(b) (color online) SRT due to the BAP mechanism with full spin-flip
electron-hole exchange scattering (solid curves) and
with only the linear  terms in the spin-flip
electron-hole exchange scattering (dotted curves) at
different electron densities
against temperature $T$.  $n_0 = 10^{11}$\ cm$^{-2}$.}
\label{fig1}
\end{figure}

We first discuss the SRT in an intrinsic GaAs QW confined by
Al$_{0.4}$Ga$_{0.6}$As barriers.
In Fig.\ \ref{fig1}(a), we plot the temperature dependence of the SRT
for a QW with well width $a=20$\ nm. The electron (heavy hole) density $n$
($p$) is $2 \times 10^{11}$\ cm$^{-2}$ and the impurity density $n_i=n$.
It is seen from the figure that the SRT due to the BAP mechanism is much larger
than that due to the DP mechanism.
Moreover, $\tau_{\mbox{\tiny BAP}}$
decreases dramatically with $T$ at low temperature,
followed by a more moderate decrease at high temperature.
The temperature dependence of $\tau_{\mbox{\tiny BAP}}$ can be
understood as follows.
When the temperature increases, more
electrons and holes tend towards the lager momentum,
hence the larger  center-of-mass momentum ${\bf K}$.
This leads to a larger the matrix element in Eq.\ (\ref{element}), and
consequently a larger scattering rate.
Furthermore, the Pauli blocking which suppresses the scattering
decreases with the increase of temperature. Both leads to the decrease of
the SRT due to the BAP mechanism.
The temperature dependence of the SRT due to the DP
mechanism has been well discussed in Refs.\
[\onlinecite{wu7,wu8,wu9,wu12}]. Therefore we will not discuss
 the DP mechanism in detail in this paper.

 \begin{figure}[bth]
\begin{center}
\includegraphics[height=5.5cm]{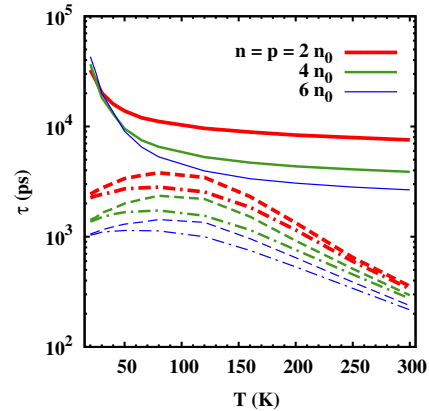}
\end{center}
\caption{(color online) SRT  due to the BAP (solid curves) and
DP (dashed curves) mechanisms and
the total SRT (dash-dotted curves) {\em vs.} temperature $T$
in intrinsic QWs
at different densities ($n =p=2$, 4, $6n_0$)
when $a = 20$\ nm and $n_i = n$. $n_0 = 10^{11}$\ cm$^{-2}$.}
\label{fig2}
\end{figure}

In order to see the difference of the SRT due to the BAP mechanism
calculated from the full spin-flip scattering [Eq.\ (\ref{BAPscat})] and the
one from the Fermi golden rule [Eq.\ (\ref{tauBAP})], i.e., neglecting the
nonlinear terms in Eq.\ (\ref{BAPscat}), we  plot the BAP SRT
calculated from the Bloch equations with only the linear terms in the
spin-flip scattering as dotted curves for
two different electron (hole) densities
in Fig.\ \ref{fig1}(b). It is noted that for high electron
density,  the SRT due to the BAP mechanism
from the Fermi golden rule is much smaller
than $\tau_{\mbox{\tiny BAP}}$ at low temperature.
Furthermore, the lower the temperature and/or the larger the electron
density, the larger the difference is due to the
``{\em breakdown}'' of the elastic scattering
approximation at low temperature and/or high density.
This is in good agreement with the condition for the
 elastic scattering.
The difference can be very small when the electron density is smaller than
$5\times 10^{10}$\ cm$^{-2}$ according our calculation.
Consequently the SRT for high electron density obtained in Ref.\ [\onlinecite{maialle}] at zero
temperature is much smaller than the actual one.
Therefore, the effect of the BAP mechanism for high electron density
at very low temperature is smaller than that claimed by
Maialle {\em et al.}
In fact, it can even be ignored. We further stress that
{\em  the effect of the
BAP mechanism at low temperature and high electron density
is far exaggerated in the literature}
due to the neglection of the nonlinear terms in the spin-flip electron-hole
exchange scattering.

In addition, in the presence of inhomogeneous broadening, any
scattering can give rise to spin R/D.\cite{wu1,wu7,wu8,lue,wu}
It is intuitive that the SRTs should satisfy:
\begin{equation}
\frac{1}{\tau_{\mbox{\tiny total}}}=\frac{1}{\tau^{\prime}_{\mbox{\tiny
      DP}}}+\frac{1}{\tau_{\mbox{\tiny BAP}}}=\frac{1}{\tau_{\mbox{\tiny DP}}}
      +\frac{1}{\tau_{\mbox{\tiny BAP}}}+\frac{1}{\tau_{\mbox{\tiny
      differ}}}\ ,
\end{equation}
where $\tau_{\mbox{\tiny BAP}}$ is directly caused by the
 spin-flip electron-hole
exchange interaction, $\tau_{\mbox{\tiny DP}}$ is from the
inhomogeneous broadening when there is no spin-flip  electron-hole
exchange interaction,
and $\tau^{\prime}_{\mbox{\tiny DP}}$ corresponds to case with the presence of
the spin-flip electron-hole exchange scattering.
The difference between $\frac{1}{\tau_{\mbox{\tiny
    DP}}}$ and $\frac{1}{\tau^{\prime}_{\mbox{\tiny DP}}}$ is noted as
$\frac{1}{\tau_{\mbox{\tiny differ}}}$.
In our calculation we found $\frac{1}{\tau_{\mbox{\tiny differ}}}$ is so small
that can be totally ignored. This is because the spin-flip electron-hole
scattering is much smaller than the other scatterings.

\begin{figure}[bth]
\begin{center}
\includegraphics[height=5.5cm]{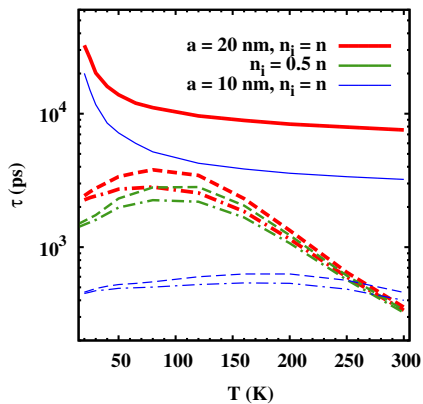}
\end{center}
\caption{(color online) SRT  due to the BAP (solid curves) and
DP (dashed curves) mechanisms and
the total SRT (dash-dotted curves) {\em vs.} temperature $T$
in intrinsic QWs
for different well widths ($a = 10$ and 20\ nm).
 $n=p=2 n_0$ and impurity densities ($n_{i} = 0.5 n$ and $n$).
 Note that the solid curves
 with the same well width but different impurity densities
 exactly coincides with each other. $n_0 = 10^{11}$\ cm$^{-2}$.}
\label{fig3}
\end{figure}

Then, we discuss the temperature dependence for different electron
densities in intrinsic QWs in Fig.\ \ref{fig2}. One can see that
$\tau_{\mbox{\tiny BAP}}$ decreases with increasing densites at high
temperature but it behaves oppositely at low temperature. On the other hand,
$\tau_{\mbox{\tiny DP}}$ decreases with increasing densities at all
temperatures. We again interpret the density
dependence of BAP mechanism by using the previous arguments:
at low temperature regime, i.e., in the degenerate limit,
the  Pauli blocking is enhanced by
increasing the carrier density and/or lowering the temperature.
Therefore, the scattering can be
suppressed by increasing density.
This causes an increase
of $\tau_{\mbox{\tiny BAP}}$. At high temperature regime,  i.e. in the
nondegenerate case, higher momentum states are occupied for larger
density. This leads to a stronger scattering and hence
 $\tau_{\mbox{\tiny BAP}}$ decreases with electron density.
From this, we find that the relative
importance of the DP and the BAP mechanisms does not change so much by changing
the electron density.

In Fig.\ \ref{fig3}, we plot the temperature dependence of the
SRTs in intrinsic QWs for different impurity
densities and well widths. It is clear that
$\tau_{\mbox{\tiny BAP}}$  does not depend on impurity
density, in other words, the curves corresponding to different
impurities concentrations exactly coincide.
However, $\tau_{\mbox{\tiny DP}}$ can
be enhanced due to the increased impurity scattering strength.
If we enlarge the well width, both  $\tau_{\mbox{\tiny DP}}$ and
$\tau_{\mbox{\tiny BAP}}$ become larger.
This is because the electron-hole exchange strength is weakened
by the form factor Eq.\ (\ref{form1}) in the scattering
  matrix elements in the BAP mechanism for wider QWs.
The leading term (linear term) of the Dresselhause
spin-orbit coupling in Eqs.\ (\ref{omegax}-\ref{omegaz}) is smaller for
wider QWs in the DP mechanism. The variation of
$\tau_{\mbox{\tiny DP}}$ is larger than $\tau_{\mbox{\tiny BAP}}$ that
is to say the relative influence of the BAP mechanism becomes more
important for wider QWs.

\begin{figure}[bth]
\begin{center}
\includegraphics[height=5.5cm]{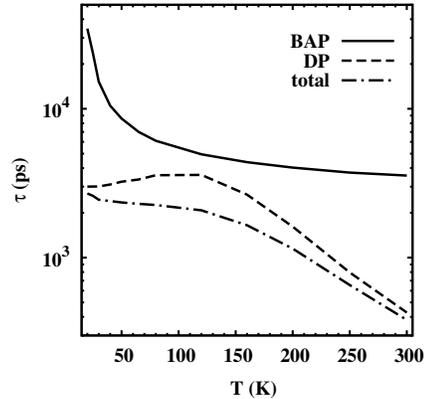}
\end{center}
\caption{SRT  due to the BAP (solid curve) and
DP (dashed curve) mechanisms and
the total SRT (dash-dotted curve) {\em vs.} temperature $T$
in $p$-type QW when $a = 20$\ nm, $n = 0.5 n_0$,
 $p_0 = 4 n_0$, and $n_i = n$. $n_0 = 10^{11}$\ cm$^{-2}$.}
\label{fig4}
\end{figure}

From our detailed investigations, we conclude that $\tau_{\mbox{\tiny BAP}}$ in
intrinsic GaAs QWs is always larger than $\tau_{\mbox{\tiny DP}}$. At
very low temperatures, the BAP mechanism can be  ignored.
However, it should
be considered at higher temperatures for accurate calculating.
Moreover, the relative importance of the BAP
mechanism is increased by raising the impurity density and the well width.

\begin{figure}[bth]
\begin{center}
\includegraphics[height=5.5cm]{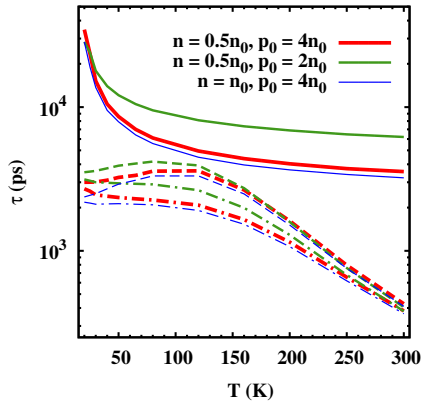}
\end{center}
\caption{(color online) SRT  due to the BAP (solid curves) and
DP (dashed curves) mechanisms and
the total SRT (dash-dotted curves) {\em vs.} temperature $T$
in $p$-type QWs with $a = 20$\ nm at different
electron densities ($n = 0.5$ and $1n_0$) and hole
 densities $p_0 = 2$ and $4n_0$. $n_i=n$. $n_0 = 10^{11}$\ cm$^{-2}$.}
\label{fig5}
\end{figure}

We now turn to study the SRT in $p$-type QWs. In Fig.\ \ref{fig4}, we
choose the well width $a=20$\ nm, $n=0.5 \times10^{11}$\ cm$^{-2}$,
$p = n + p_0=n+4 \times10^{11}$\ cm$^{-2}$, and $n_i=n$.
One can see that the magnitudes of $\tau_{\mbox{\tiny DP}}$
and $\tau_{\mbox{\tiny
BAP}}$ are very  close around $T = 150$\ K.  In $p$-type
QWs, both the spin-conserving
and spin-flip electron-hole scatterings  are
greatly enhenced by increasing the hole density.
The former gives rise to the
increase of $\tau_{\mbox{\tiny DP}}$ in the strong scattering
limit\cite{wu12,lue} and the latter gives rise to the
decrease of $\tau_{\mbox{\tiny BAP}}$. Therefore both SRTs are
getting closer for larger hole concentration.
In the case of Fig.\ \ref{fig4}, the contributions from the
DP and BAP mechanisms are nearly the same around $150$\ K, and at
lower and higher temperatures, the contribution from the DP mechanism is
no more than one order of magnitude larger than the BAP one.
In addition, $1/\tau_{\mbox{\tiny differ}}$ is
still very small and can be totally ignored.

\begin{figure}[bth]
\begin{center}
\includegraphics[height=5.5cm]{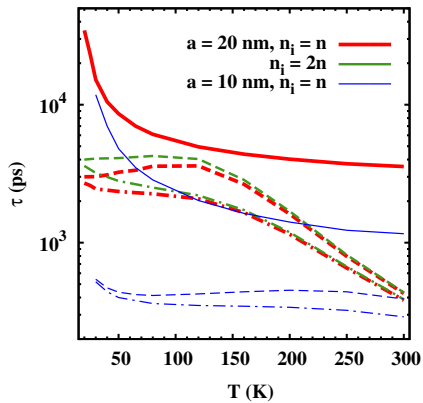}
\end{center}
\caption{(color online) SRT  due to the BAP (solid curves) and
DP (dashed curves) mechanisms and
the total SRT (dash-dotted curves) {\em vs.} temperature $T$
in $p$-type QWs with $n=0.5 n_0$,
  $p_0 = 4 n_0$ at different well widths ($a = 10$ and 20\ nm) and
  impurity densities ($n_{i} = n$ and $2n$). $n_0 = 10^{11}$\ cm$^{-2}$.}
\label{fig6}
\end{figure}

We now analyze the temperature dependence of the
SRT for different electron and hole densities in $p$-type QWs. In
Fig.\ \ref{fig5}, the calculated SRT for different electron and hole
densities are shown. In Fig.\ \ref{fig6}, a similar analysis is made
for different well widths and impurity densities.
The general features can be understood from the following. When
the electron density becomes larger, both $\tau_{\mbox{\tiny DP}}$ and
$\tau_{\mbox{\tiny BAP}}$ become smaller with similar amplitude. (Note
that $n=0.5 n_0$ and  $n_0$ are both within the nondegenerate limit.) When
hole density gets larger, both $\tau_{\mbox{\tiny DP}}$ and
$\tau_{\mbox{\tiny BAP}}$ become smaller with the amplitude of the
latter being larger than the former (i.e., the importance of the BAP
mechanism gets increased).
This is because the electron-heavy hole scattering is markedly enhanced
with the hole density.
As the BAP mechanism is determined by the hole density,
$\tau_{\mbox{\tiny BAP}}$ is very sensitive to the hole density.
Nevertheless, $\tau_{\mbox{\tiny DP}}$ is less sensitive as
it is also determined by all the other scatterings.
When the well width gets larger,
$\tau_{\mbox{\tiny DP}}$ is enhanced with a larger amplitude at low
temperature and with a small amplitude at high temperature, whereas
$\tau_{\mbox{\tiny BAP}}$ becomes larger moderately.
These results are similar to Fig. \ref{fig3}.
Consequently the BAP mechanism
becomes more important, especially around $T= 150$\ K in the
present case. When the impurity density gets
larger, $\tau_{\mbox{\tiny DP}}$ becomes larger and $\tau_{\mbox{\tiny
BAP}}$ does not change. This makes the relative effect of
BAP mechanism become larger.

From above features, we emphasize that the BAP mechanism is
important in  $p$-type QWs, especially for large well width and/or large hole
densities (i.e. heavily doped) and large impurity densities.
It is very different from the bulk systems in which the BAP mechanism
is absolutely dominant at low temperature.
Therefore, both the BAP and the DP mechanisms should be
considered to get the right SRT in QWs.

\section{SUMMARY}

In summary, we have investigated the SRT due to both the DP and BAP mechanisms
in intrinsic and $p$-type GaAs (001) QWs
by constructing and numerically solving the fully
microscopic kinetic spin Bloch equations.
We consider all the relevant
scatterings such as the electron-AC phonon, electron-LO phonon,
electron-nonmagnetic impurity, and  electron-electron Coulomb
scattering. Furthermore, the spin-conserving electron-heavy hole
 scattering, which enhances the total
scattering strength and therefore $\tau_{\mbox{\tiny DP}}$, and
the spin-flip electron-hole exchange scattering,
which induces the BAP SRT, are also included.

We stress it is very important to
calculate the SRT from our fully microscopic
approach, especially at high electron density and
 low temperatures where the nonlinear terms in the
electron-hole exchange scattering becomes very important. The SRT
obtained from our fully microscopic approach is much larger than
that from the Fermi golden rule. This means that {\em the BAP mechanism is
negligible at very low temperature and high electron density.}
We speculate  this is also true in the bulk case.
This is very different from the
predictions in the literature.

We investigate the temperature dependence of the
SRTs: The SRT due to the BAP mechanism   $\tau_{\mbox{\tiny BAP}}$
 decreases rapidly with increasing temperature
at very low temperature and slowly
at higher temperature for both intrinsic and
$p$-type QWs.
It also decreases with electron density for both intrinsic and
$p$-type QWs. For $p$-type semiconductors,
it further decreases with hole density.
We also compare the relative importance of the SRTs from the
BAP and DP mechanisms. The SRT from the DP mechanism is
also calculated from the kinetic spin Bloch equations which
give the SRT also quite different from that from the single-particle
approach as discussed extensively in our previous
works.\cite{wu,wu7,wu8,wu12,wu11}
We find in intrinsic QWs, the effect of the
BAP mechanism is much smaller than that from the DP mechanism
at low temperature and it is smaller by nearly one order of
magnitude at higher temperature;
In $p$-type QWs, the SRT from the BAP mechanism is comparable
with the one from the  DP mechanism around certain temperature
(such as $150$\ K in the case we study), especially when the hole
density and/or the width of the QWs are large. For both the
intrinsic and $p$-type QWs, the contribution from the BAP mechanism
at very low temperature are negligible.
We conclude that the spin R/D in QWs
is very different from the bulk samples. In 2D case
the BAP mechanism hardly dominates the spin relaxation.
 Instead, it is either smaller or comparable to the DP mechanism.

\begin{acknowledgments}

This work was supported by the Natural
Science Foundation of China under Grant Nos.\ 10574120 and
10725417, the
National Basic Research Program of China under Grant
No.\ 2006CB922005 and the Knowledge Innovation Project of Chinese Academy
of Sciences. The authors would like to thank  Dan Csontos
for critical reading of this manuscript and C. L\"u
for helpful discussions.
\end{acknowledgments}

\begin{appendix}

\begin{section}{Coherent and spin-conserving  scattering terms
in kinetic spin Bloch equations}

The coherent terms can be written as
\begin{widetext}
\begin{eqnarray}
\left.\frac{\partial f_{{\bf k},\sigma}}{\partial t}\right|_{\mbox{\small
coh}}&=&-\sigma\Bigl[\Omega_x({\bf k})\mbox{Im}\rho_{\bf k}+\Omega_y({\bf
  k})\mbox{Re}\rho_{\bf k}\Bigr]+2\sigma\mbox{Im}\sum_{\bf
  q}V_{ee,q}\rho_{\bf k+q}^{\ast}\rho_{\bf k}\ , \\
\left. \frac{\partial \rho_{\bf k}}{\partial t}\right |_{\mbox{\small
coh}}&=&\frac{1}{2}\Bigl[i \Omega_x({\bf k})+\Omega_y({\bf k})\Bigr](f_{{\bf
    k},+1}-f_{{\bf k},-1})+i\sum_{\bf q}V_{ee,q}\Bigl[(f_{{\bf
    k+q},+1}-f_{{\bf k+q},-1})\rho_{\bf k}-\rho_{\bf k+q}(f_{{\bf
    k},+1}-f_{{\bf k},-1})\Bigr]\ ,
\end{eqnarray}
where $V_{ee,q}=\frac{\sum_{q_z}v_{Q}f_{e}(q_z)}{\epsilon ({\bf q})}$,

The electron-impurity scattering terms read
\begin{eqnarray}
\left.\frac{\partial f_{{\bf k},\sigma}}{\partial t}\right|_{\mbox{\small
im}}&=&\Bigl\{-2\pi n_{i}\sum_{\bf q}U^2_{q}\delta(\varepsilon^e_{\bf
  k}-\varepsilon^e_{\bf k-q})\Bigl[f_{{\bf k},\sigma}(1-f_{{\bf
      k-q},\sigma})-\mbox{Re}(\rho_{\bf k} \rho^{\ast}_{\bf k-q})
  \Bigr]\Big\}-\Bigl\{{\bf k}\leftrightarrow {\bf k-q}\Bigr\}\ , \\
\left. \frac{\partial \rho_{\bf k}}{\partial t}\right |_{\mbox{\small
im}}&=&\Bigl\{\pi n_{i}\sum_{\bf q}U^2_{q}\delta(\varepsilon^e_{\bf
  k}-\varepsilon^e_{\bf k-q})\Bigl[(f_{{\bf k},+1}+f_{{\bf
      k},-1})\rho_{\bf k-q}-(2-f_{{\bf k-q},+1}-f_{{\bf
      k-q},-1})\rho_{\bf k}\Bigr]\Bigr\}-\Bigl\{{\bf k}\leftrightarrow {\bf
  k-q}\Bigr\}\ ,
\end{eqnarray}
\end{widetext}
in which $\Bigl\{{\bf k}\leftrightarrow {\bf k-q}\Bigr\}$ stands for
the same terms previously in $\Bigl\{ \Bigr\}$ but interchanging ${\bf k}\leftrightarrow {\bf k-q}$.
In these equations $U^2_{\bf q}=\sum_{q_z}(Z_{i}v_{Q}/\epsilon({\bf
  q}))^2f_{e}(q_z)$ with $Z_{i}$ (assumed to be 1 in our calculation)
  the charge number of the impurity.
The electron-phonon scattering terms are
\begin{widetext}
\begin{eqnarray}
\left.\frac{\partial f_{{\bf k},\sigma}}{\partial t}\right|_{\mbox{\small
ph}}&=&\Bigl\{-2\pi\sum _{{\bf q}q_{z},\lambda}
g^2_{{\bf q} q_{z},\lambda}\delta
(\varepsilon^e_{\bf k}-\varepsilon^e_{\bf k-q}-
\Omega_{{\bf q} q_{z},\lambda})  %\nonumber\\
%&&\hspace{0.6cm}\times
[N_{{\bf q} q_{z},\lambda}
(f_{{\bf k},\sigma}-f_{{\bf k-q},\sigma})
+f_{{\bf k},\sigma}(1-f_{{\bf k-q},\sigma})\nonumber\\
&&-\mbox{Re}(\rho_{\bf k} \rho^{\ast}_{\bf k-q})]\Bigr\}
-\Bigl\{{\bf k} \leftrightarrow {\bf k-q}\Bigr\}\ ,\\
\left. \frac{\partial \rho_{\bf k}}{\partial t}\right |_{\mbox{\small
ph}}&=&\Bigl\{
  \pi\sum_{{\bf q}q_z,\lambda}g^2_{{\bf q}q_z,\lambda}
  \delta(\varepsilon^e_{\bf k}-\varepsilon^e_{\bf k-q}
  -\Omega_{{\bf q}q_z,\lambda})
[\rho_{\bf k-q}(f_{{\bf k},+1}+f_{{\bf k},-1})
  +(f_{{\bf k-q},+1}+f_{{\bf k-q},-1}-2)\rho_{\bf k}\nonumber\\
&&-2N_{{\bf q} q_{z},\lambda}
  (\rho_{\bf k}-\rho_{\bf k-q})]
\Bigr\} -\Bigl\{{\bf k}\leftrightarrow {\bf k-q}\Bigr\}\ ,
\end{eqnarray}
\end{widetext}
where $\lambda$ represents the phonon mode. For the
electron--longitudinal-optic-phonon (LO) scattering,
the matrix element $g^2_{{\bf Q},\mbox{LO}}=\{2\pi^2\Omega_{\mbox{\tiny
    LO}}/[(q^2+q_z^2)]\}(\kappa_{\infty}^{-1}-\kappa_0^{-1})f_e(q_z)$;
for electron--acoustic-phonon scattering due to the deformation potential,
$g^2_{{\bf Q},def}=\frac{\hbar\Xi^{2} Q}{2dv_{sl}}f_e(q_z)$;
and for that due to the piezoelectric coupling,
$g^2_{{\bf Q},pl}=\frac{32\pi^{2}\hbar e^{2} e_{14}^{2}}{\kappa_0^{2}}
\frac{(3q_{x}q_{y}q_{z})^2}{dv_{sl}Q^7}f_e(q_z)$ for the longitudinal
phonon and $g^2_{{\bf Q},pt}=\frac{32\pi^{2}\hbar e^{2} e_{14}^{2}}
{\kappa_0^{2}}\frac{1}{dv_{st}Q^{5}}
(q_{x}^{2}q_{y}^{2}+q_{y}^{2}q_{z}^{2}+q_{z}^{2}q_{x}^{2}
-\frac{(3q_{x}q_{y}q_{z})^2}{Q^2})f_e(q_z)$ for the transverse
phonon. Here $\Xi=8.5$\ eV is the deformation potential;
$d=5.31$\ g/cm$^3$ is the mass
density of the crystal; $v_{sl}=5.29\times 10^3$\ m/s ($v_{st}
=2.48\times 10^3$\ m/s) is the velocity of the
longitudinal (transverse) sound wave; $\kappa_0=12.9$ denotes the
static dielectric constant and $\kappa_{\infty}=10.8$ denotes the
optical dielectric constant;
and $e_{14}=1.41\times 10^9$\ V/m represents the piezoelectric
constant.
$\Omega_{\mbox{\tiny LO}}=35.4$\ meV is the LO phonon frequency, and the
AC phonon spectra $\Omega_{{\bf Q}\lambda}$ are given by
$\Omega_{{\bf Q}l}=v_{sl}Q$ for the longitudinal mode and
$\Omega_{{\bf Q}t}=v_{st}Q$ for the transverse mode.\cite{parameter}
$N_{{\bf q} q_{z},\lambda}=[\mbox{exp}(\beta \Omega_{{\bf q} q_{z},\lambda})-1]^{-1}$ represents the Bose distribution.

The spin-conserving electron-electron Coulomb scattering terms are given by
\begin{widetext}
\begin{eqnarray}
\left.\frac{\partial f_{{\bf k},\sigma}}{\partial t}\right|_{\mbox{\small
ee}}&=&\Bigl\{-2\pi\sum_{{\bf q,k^{\prime}},\sigma^{\prime}}V^2_{ee,q}\delta
(\varepsilon^e_{\bf k-q}-\varepsilon^e_{\bf k}+\varepsilon^e_{\bf
  k^{\prime}}-\varepsilon^e_{\bf k^{\prime}-q}) \Bigl[(1-f_{{\bf
      k-q},\sigma})f_{{\bf k},\sigma}(1-f_{{\bf
      k^{\prime}},\sigma^{\prime}})f_{{\bf
      k^{\prime}-q},\sigma^{\prime}}\nonumber \\
&+&\frac{1}{2}\rho_{\bf k} \rho^{\ast}_{\bf k-q}(f_{{\bf
      k^{\prime}},\sigma^{\prime}}-f_{{\bf
      k^{\prime}-q},\sigma^{\prime}})+\frac{1}{2}\rho_{\bf k^{\prime}}
  \rho^{\ast}_{\bf k^{\prime}-q}(f_{{\bf k-q},\sigma}-f_{{\bf
      k},\sigma})\Bigr] \Bigr\}-\Bigl\{{\bf k}\leftrightarrow {\bf k-q},
      {\bf k^{\prime}}\leftrightarrow {\bf k^{\prime}-q}\Bigr\}\ ,\\
\left. \frac{\partial \rho_{\bf k}}{\partial t}\right |_{\mbox{\small
ee}}&=&\Bigl\{-\pi\sum_{{\bf q,k^{\prime}},\sigma^{\prime}} V^2_{ee,q}\delta
(\varepsilon^e_{\bf k-q}-\varepsilon^e_{\bf k}+\varepsilon^e_{\bf
  k^{\prime}}-\varepsilon^e_{\bf k^{\prime}-q})\Bigl[(f_{{\bf
      k-q},+1}\rho_{\bf k}+f_{{\bf k},-1}\rho_{\bf k-q})(f_{{\bf
      k^{\prime}},\sigma^{\prime}}-f_{{\bf
      k^{\prime}-q},\sigma^{\prime}})\nonumber \\
&+&\rho_{\bf k}[(1-f_{{\bf k^{\prime}},\sigma^{\prime}})f_{{\bf
      k^{\prime}-q},\sigma^{\prime}}-\mbox{Re}(\rho_{\bf k^{\prime}}
    \rho^{\ast}_{\bf k^{\prime}-q})]
-\rho_{\bf k-q}[f_{{\bf k^{\prime}},\sigma^{\prime}}(1-f_{{\bf
      k^{\prime}-q},\sigma^{\prime}})-\mbox{Re}(\rho^{\ast}_{\bf
    k^{\prime}} \rho_{\bf k^{\prime}-q})]\Bigr]\Bigr\}\nonumber \\
&-&\Bigl\{{\bf k}\leftrightarrow {\bf k-q},
      {\bf k^{\prime}}\leftrightarrow {\bf k^{\prime}-q}\Bigr\}\ .
\end{eqnarray}
\end{widetext}
\end{section}
\end{appendix}


\begin{thebibliography}{0}
\bibitem{comment1} It is noted that the screening in the Hartree-Fock terms in
$\partial \rho_{{\bf k},\sigma\sigma^\prime}/
\partial t|_{\mbox{coh}}$ in Eq.\ (\ref{bloch})
is also updated by the current
one with the contributions from the heavy holes.
\end{thebibliography}
\end{document}